\documentclass{nature}
\usepackage{graphicx}

\title{Scale invariance and universality of force networks in static granular
	matter}

\author{Srdjan Ostojic$^{1}$, Ell\'ak Somfai$^2$ \& Bernard Nienhuis$^1$}

\begin{document}

\maketitle

\begin{affiliations}
 \item Institute for Theoretical Physics, Universiteit van Amsterdam,
    Valckenierstraat 65, 1018 XE Amsterdam, Netherlands
 \item Universiteit Leiden, Instituut-Lorentz, PO Box 9506, 2300 RA Leiden,
    Netherlands
\end{affiliations}

\begin{abstract}
Force    networks    form   the    skeleton    of   static    granular
matter\cite{pg2005,  jaeger:rev}.   They  are  the key  ingredient  to
mechanical     properties,     such    as     stability\cite{daerr99},
elasticity\cite{makse99,       resp:goldenberg05}       and      sound
transmission\cite{jia99,somfai05}, which are  of utmost importance for
civil  engineering and industrial  processing.  Previous  studies have
focused  on  the  global  structure of  external  forces\cite{vanel99,
wittmer96, resp:geng1,resp:reydellet} (the boundary condition), and on
the     probability     distribution     of     individual     contact
forces\cite{makse99,majmudar05}.  The  disordered spatial structure of
the  force network,  however, has  remained elusive  so far.   Here we
report  evidence for scale  invariance of  clusters of  particles that
interact via relatively strong  forces.  We analyzed granular packings
generated  by molecular dynamics  simulations mimicking  real granular
matter;  despite  the visual  variation,  force  networks for  various
values of  the confining pressure and other  parameters have identical
scaling  exponents   and  scaling  function,  and   thus  determine  a
universality   class.   Remarkably,   the  flat   ensemble   of  force
configurations\cite{edwards1,    jam:makse1,fluct:jacco1}---a   simple
generalization of  equilibrium statistical mechanics---belongs  to the
same  universality   class,  while  some   widely  studied  simplified
models\cite{qmodel0,qmodel, raj} do not.
\end{abstract}

\begin{figure}
\begin{center} 
\begin{minipage}{0.32\linewidth}
\includegraphics [width=0.95\linewidth]{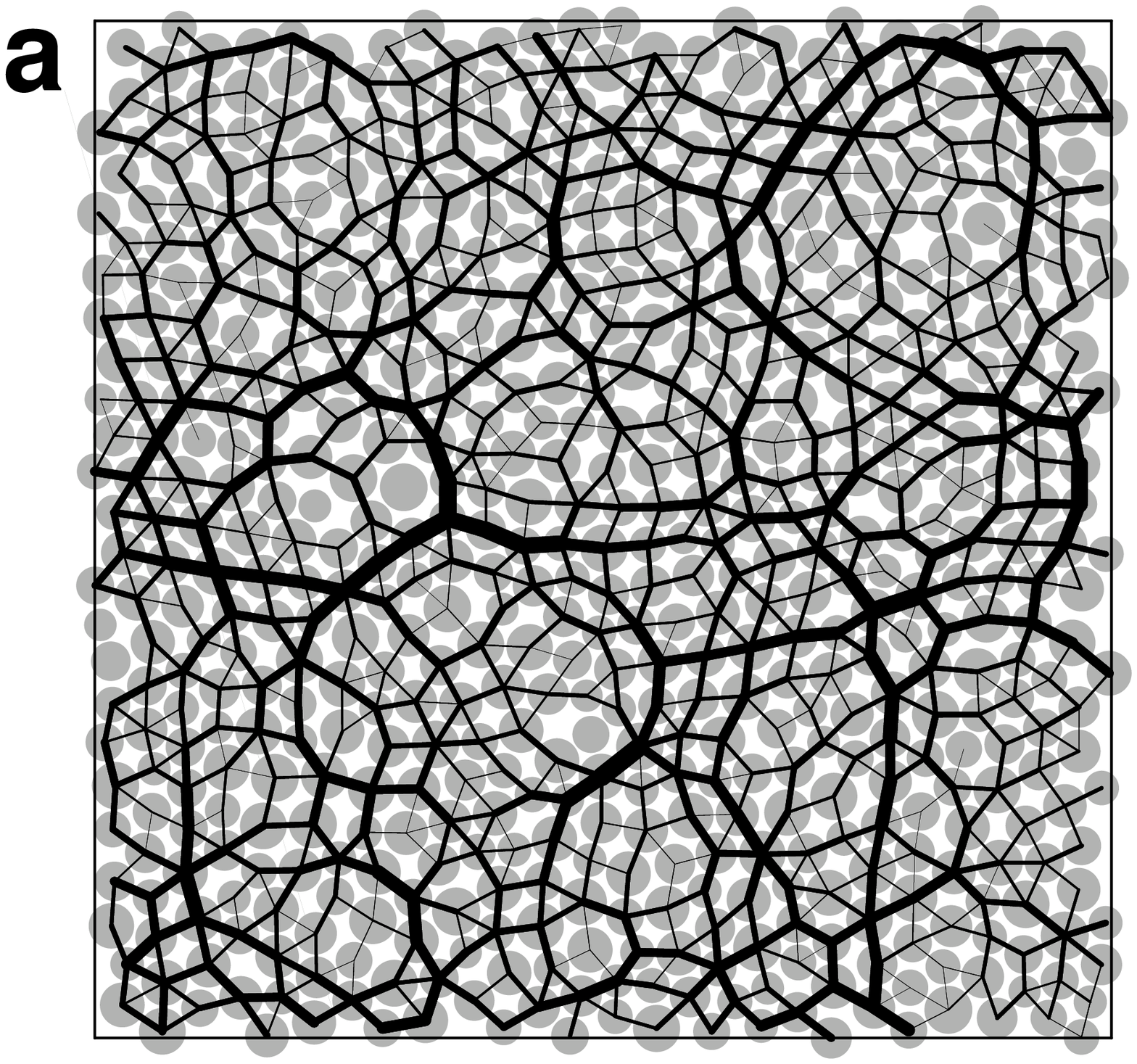}
\end{minipage}
\begin{minipage}{0.32\linewidth}
\includegraphics [width=0.95\linewidth]{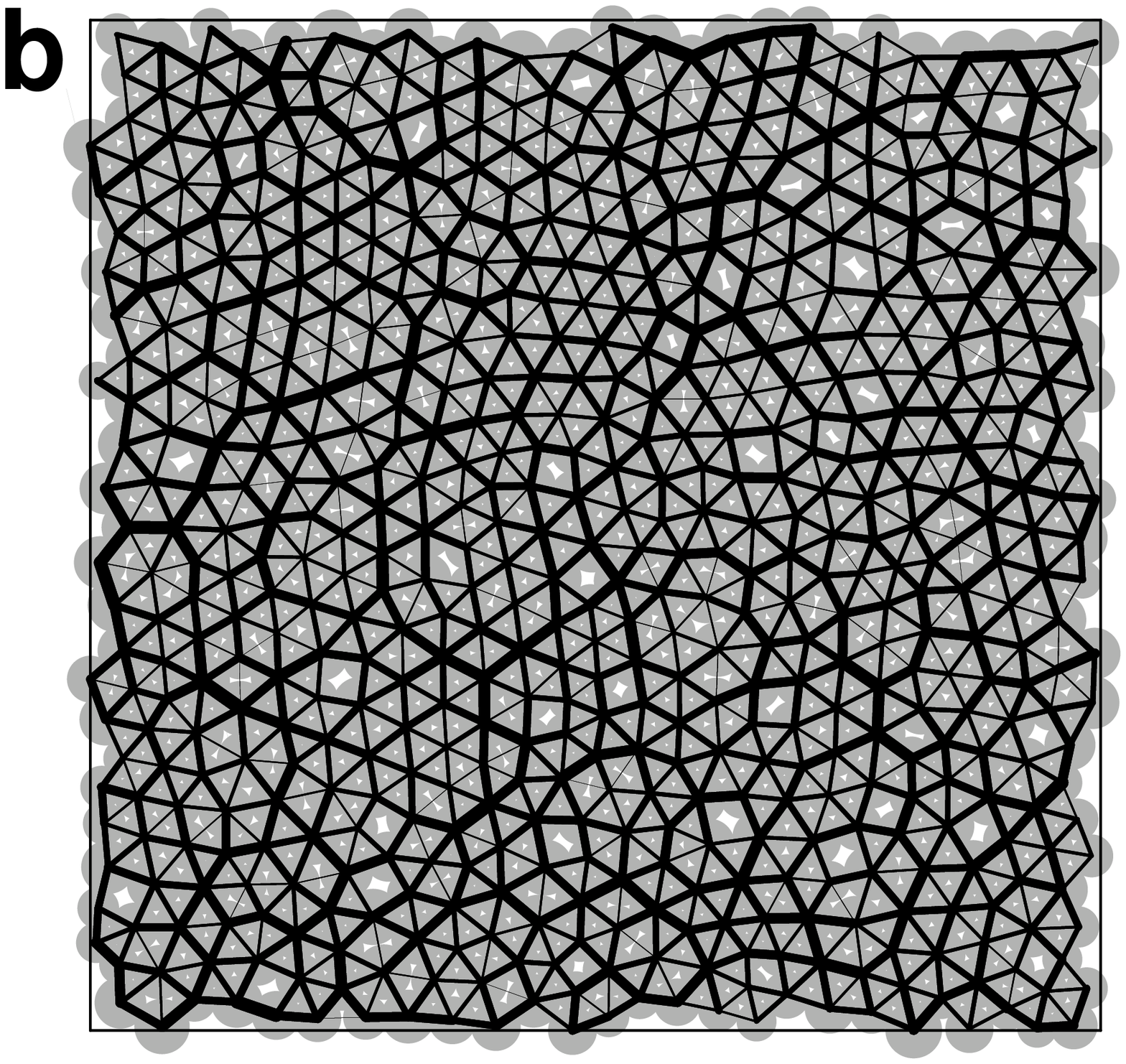}
\end{minipage}
\begin{minipage}{0.32\linewidth}
\includegraphics [width=0.95\linewidth]{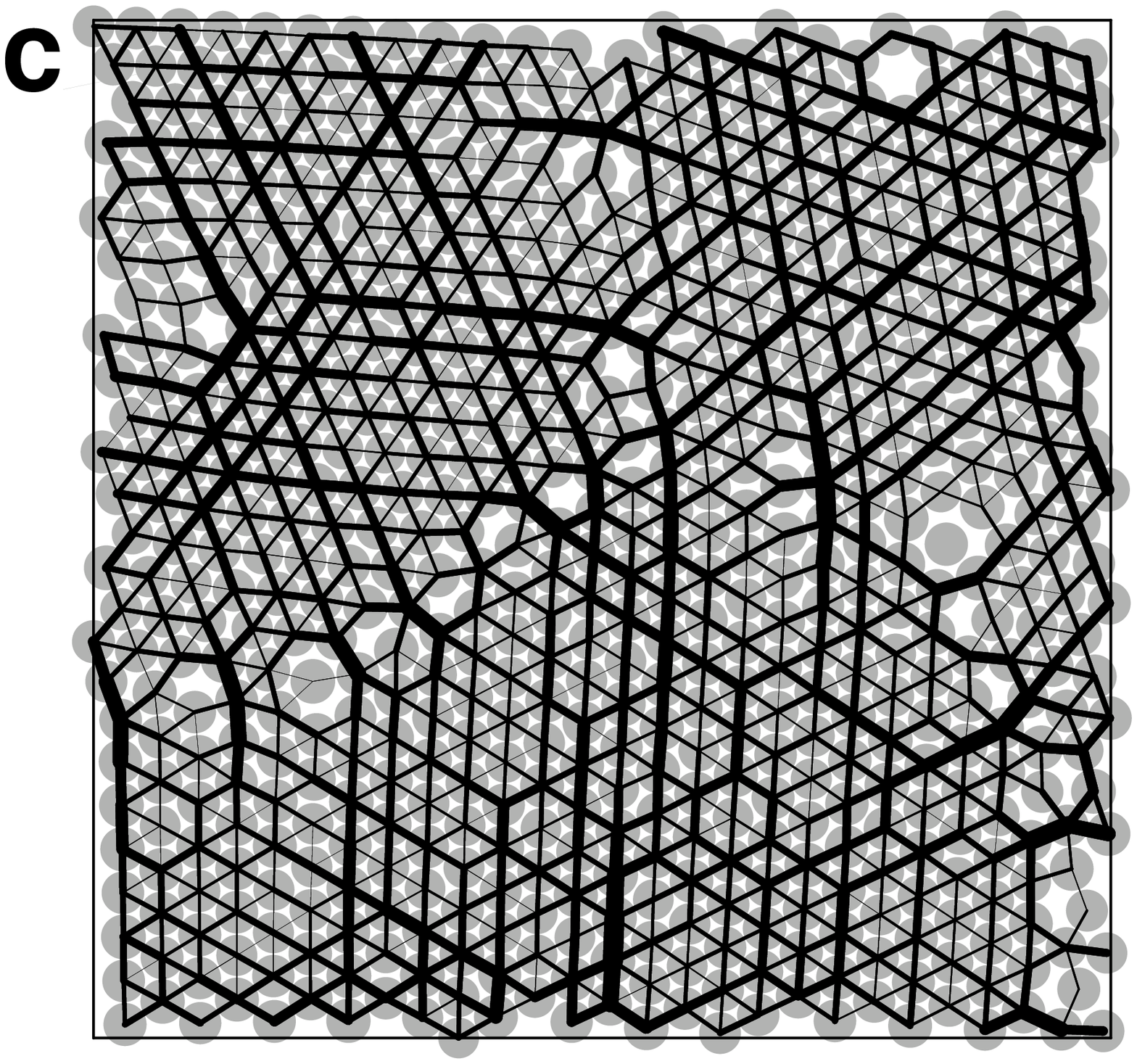}
\end{minipage}
\caption{Dependence of the force network on pressure and
  polydispersity. Grains are represented as gray disks and
  forces as bonds .  The thickness  of each bond
  is  proportional to  the magnitude of the force.
  Packing under (a) pressure $p=10^{-4}$ (in  rescaled units)  and 
  polydispersity $d=20\% $  (b)  $p=10^{-1}$ and $d=20\% $, (c) $p=10^{-2}$   and  $d=5\% $ \label{fig1} }
\end{center}
\end{figure}

Sand dunes,  piles of apples in  a supermarket, and coffee  beans in a
jar are all examples of static granular matter.  In these examples the
particles  (sand grains,  apples,  coffee beans)  interact with  their
neighbors  with repulsive contact  forces forming  a network.   Due to
disorder (even in a regular pile each apple is slightly different) the
force   network    is   inhomogeneous   in   space\cite{fluct:radjai1,
fluct:makse}.  The  simplest characterization of the  force network is
the     probability     distribution     of     individual     contact
forces\cite{fluct:mueth,  fluct:blair1,  fluct:erikson}:  $P(F)$.   In
typical granular packings $P(F)$ has an exponential tail at large $F$,
and a  plateau at small  $F$.  While $P(F)$  is an important  and well
studied quantity, it tells nothing  about the spatial structure of the
force  network, which  is  so striking  visually,  as can  be seen  on
Fig.~\ref{fig1}~a.  As a large contact force on one side of a particle
is typically  balanced by  another large force  on the  opposite side,
large  forces   tend  to  align  in   filamentary  structures,  called
\emph{force chains} though they  are not completely linear.  The force
chains are stabilized  by weaker side contacts, and  form a disordered
branching  network   resembling  a  fractal.    Typical  fractals  are
invariant under a change of  length scale.  A quantitative analysis of
that scale-invariance  is however difficult, in part  because there is
no  sharp distinction  between  the force  chains  and the  background
contacts.

In this Letter we show that discriminating between forces of different
magnitudes provides  a quantitative characterization  of force chains.
Via an analogy with equilibrium critical phenomena, such a description
naturally  captures  scale  invariance:  a set  of  scaling  exponents
(fractal dimensions) and a scaling function are extracted from a given
ensemble   of   force  networks.    We   applied   this  approach   to
two-dimensional granular  packings under isotropic  pressure generated
by  molecular dynamics  simulations.  Although  the appearance  of the
force network  varies significantly with  pressure and polydispersity,
we  found that  the scaling  exponents  and the  scaling function  are
independent of these and  other parameters, and thus capture universal
properties of force networks.  We also show that a theoretical model based
on  the extension  of  equilibrium statistical  mechanics proposed  by
Edwards  faithfully  represents this  universality  class, while  some
other simple models do not.

\begin{figure}
\begin{center} 
\begin{minipage}{0.32\linewidth}
\includegraphics [width=0.95\linewidth]{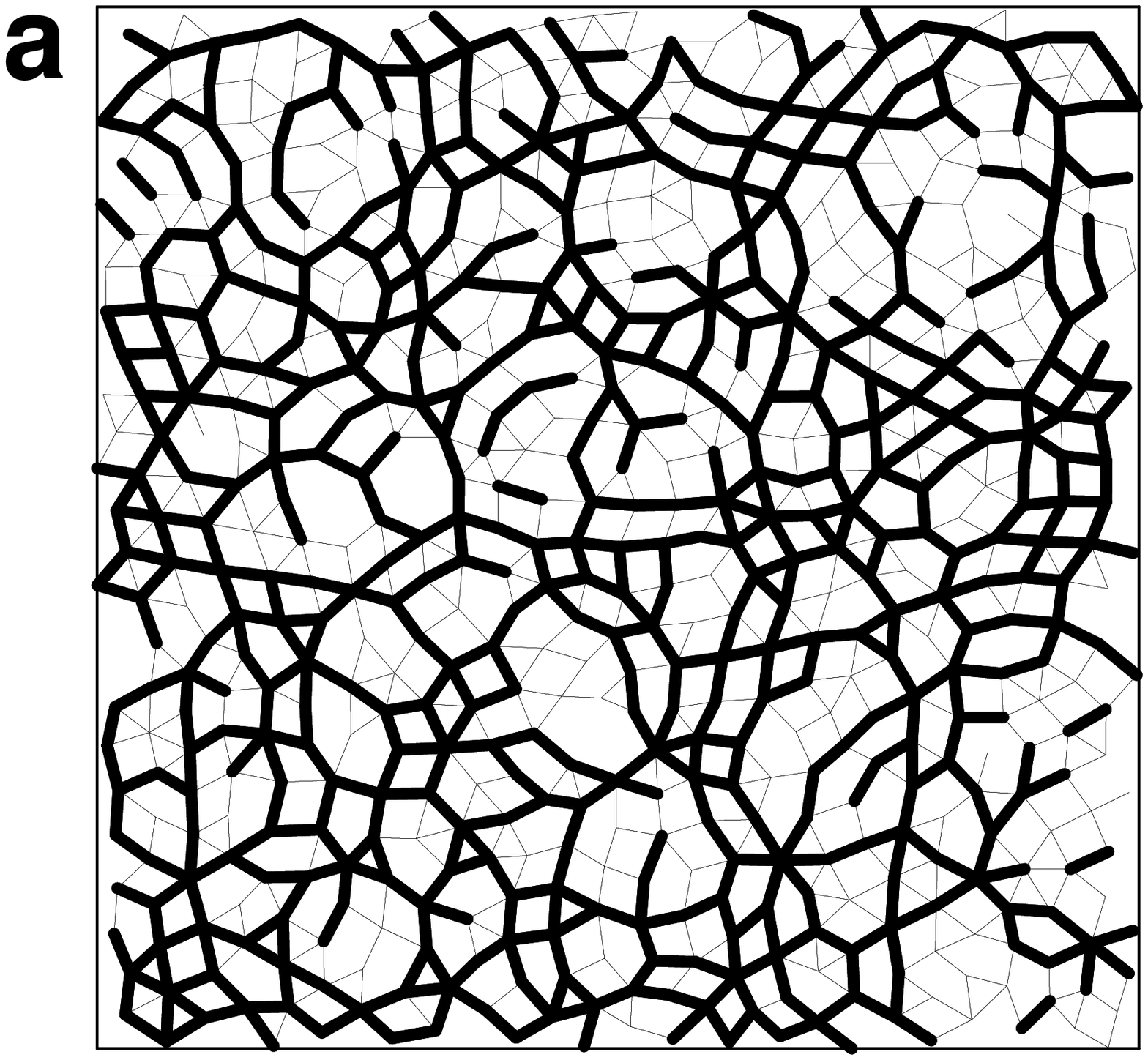}
\end{minipage}
\begin{minipage}{0.32\linewidth}
\includegraphics [width=0.95\linewidth]{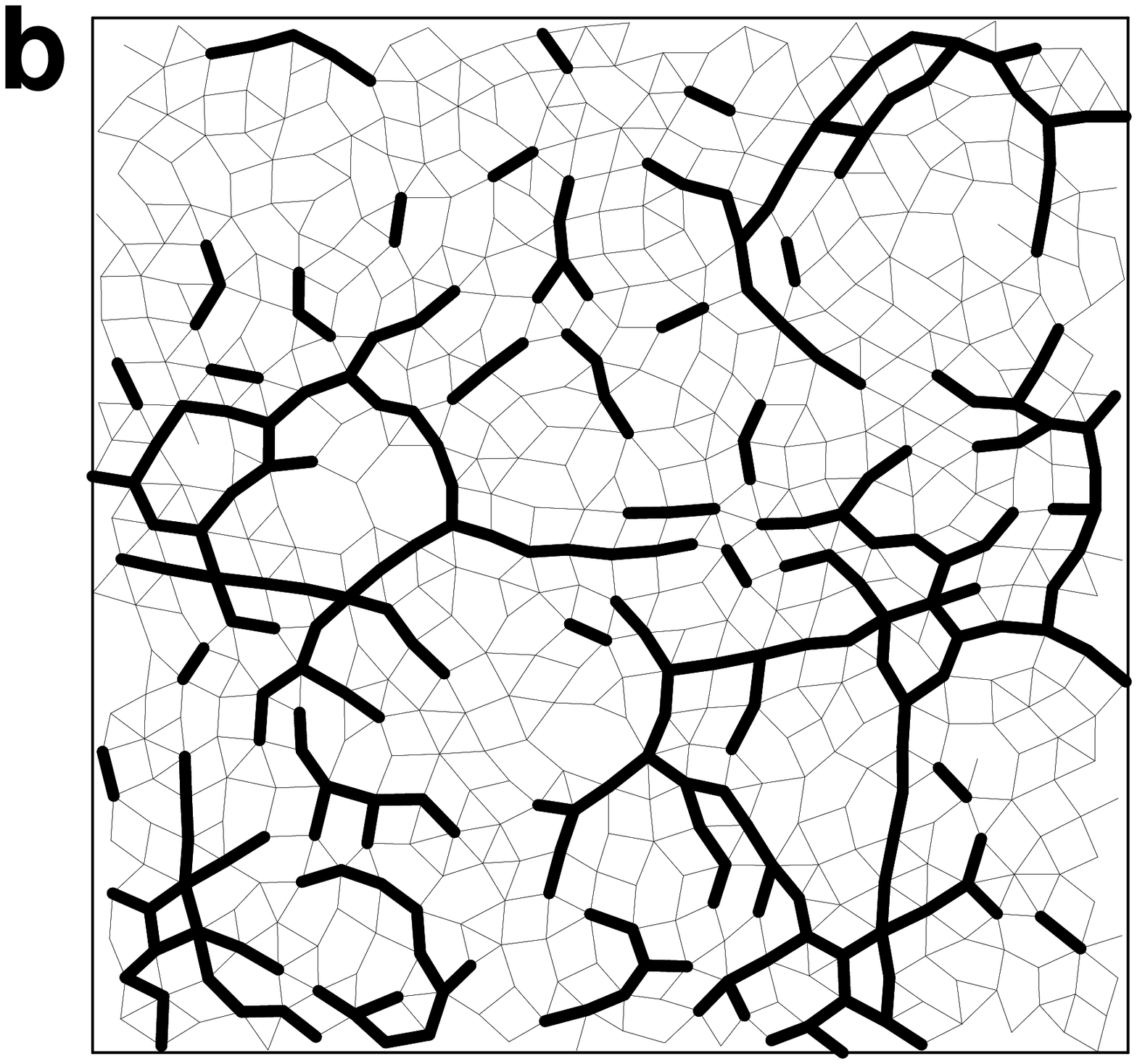}
\end{minipage}
\begin{minipage}{0.32\linewidth}
\includegraphics [width=0.95\linewidth]{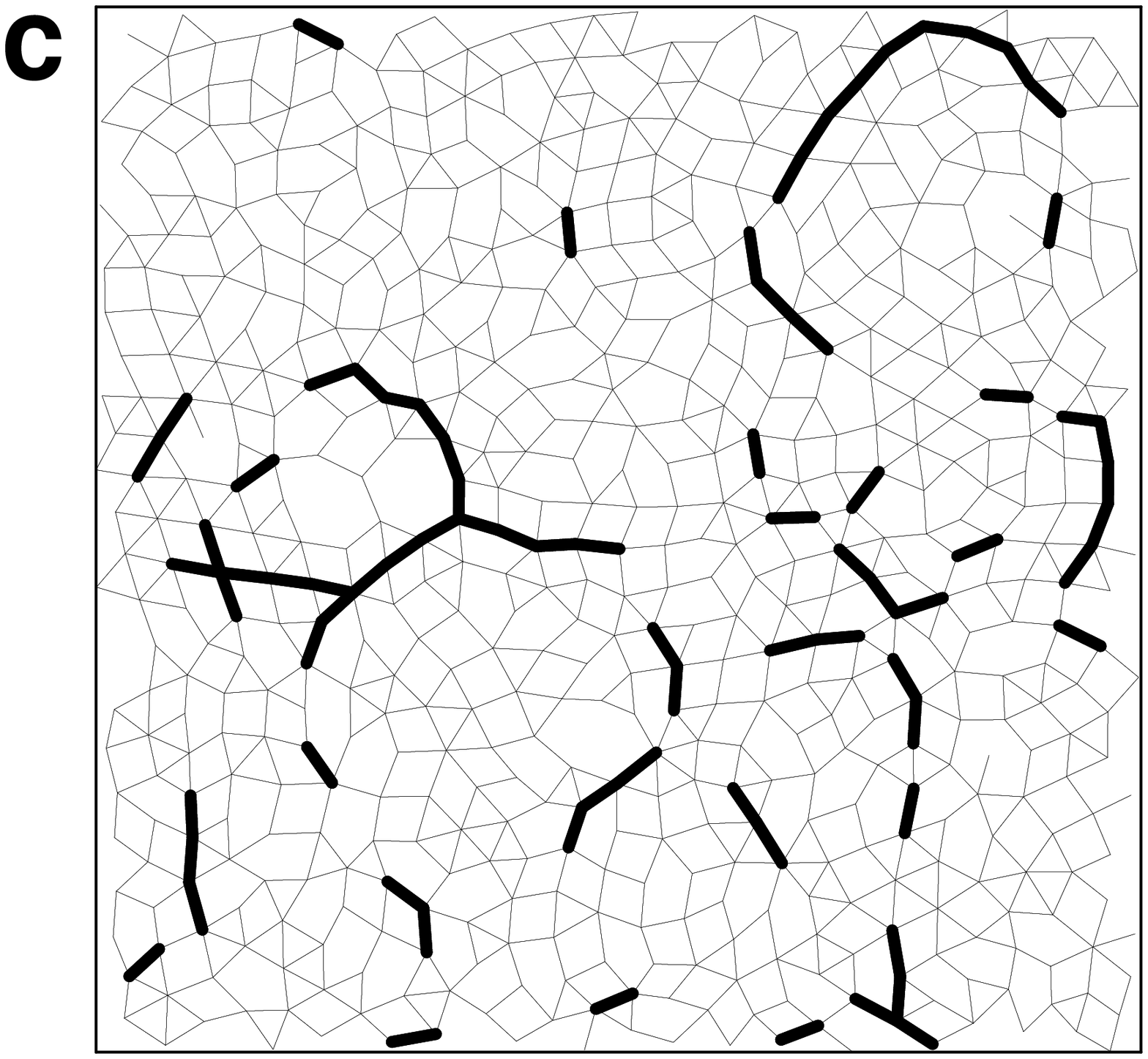}
\end{minipage}
\caption{   Force  chains   at  different   scales.  The   packing  of
  Fig.~\ref{fig1}~a  is repeated  here,  showing as  bold lines  forces
  larger  then a  threshold  $f$. The  threshold values  are (a)
  $f=0.5f_c$,  (b) $f=f_c$  and  (c) $f=1.5f_c$,  where  $f_c$ is  the
  critical threshold.\label{fig2}}
\end{center}
\end{figure}

The force network in a granular packing can be represented as a set of
bonds connecting grains in contact.   Each bond carries a scalar given
by  the  magnitude of  the  repulsive  force  between the grains  (see
Fig.~\ref{fig1}).  To quantify the, visually obvious, force chains in such
a network, we choose a  threshold $f$, and consider only chains formed
by bonds carrying forces larger  than $f$.  Rather than select a fixed
arbitrary value of  $f$, we study such structures  at different scales
by varying the threshold.  For small values, most of the grains remain
connected, but  as the  threshold is increased  the packing  breaks up
into disconnected clusters (see  Fig.~\ref{fig2}).  The extent of each
force chain can then be characterized by the size of the corresponding
cluster, i.e.  the number of mutually connected bonds.

The force  network varies totally  between packings created  under the
same external  conditions, so that  a statistical approach  is needed.
Following the above procedure, force chains in an ensemble of packings
can be described by the probability $P(s,f)$ that, at a threshold $f$,
a  given random  grain  belongs to  a  cluster of  size $s$.   Similar
statistical descriptions in terms of clusters are common in the theory
of equilibrium phase transitions \cite{fort-kast}, where the threshold
$f$  plays  the  role  of  temperature.  As  in  a  percolation  model
\cite{stauffer}, a  phase transition occurs at  the critical threshold
$f_c$ above which  no cluster connects the opposite  boundaries of the
system.  Around  this value, the system exhibits  scale invariance, in
particular  the  probability  distribution  of cluster  sizes  can  be
expressed  as  $P(s,f)\sim  s^{-\tau}\rho  (s/(f-f_c)^\sigma)$.   This
scale invariance is {\it  universal}: the scaling exponents $\tau$ and
$\sigma$, as well  as the scaling function $\rho$  are determined only
by the  global symmetries  of the system  rather than  the microscopic
details  of  the  interactions.   Equilibrium critical  phenomena  can
therefore  be  classified  in  a  discrete number  of  {  universality
classes} according to the values of the exponents.

Such  an  analogy with  equilibrium  critical  phenomena suggests  the
existence of  scale invariance around the threshold  $f_c$ below which
one of the clusters spans  the entire packing.  The associated scaling
exponents and scaling function  would provide a novel characterization
of the  spatial organization of  forces in granular matter.   To study
the  existence of scale  invariance and  universality, we  examine the
behavior of the  $n^{th}$ moment $m_n$ of $P(s,f)$  as function of the
threshold  $f$ and  the system  size $N$  (number of  contacts  in the
packing). We look for scaling as function of the system size, which in
case of equilibrium critical phenomena takes the form
\begin{equation}
m_n(f,N)\sim N^{\phi_n}{M_n}\left( [f-f_c]{N}^{1/2\nu}\right),
\label{fss}
\end{equation}
 where the
scaling function $M_n$ is related via integration to $\rho$, and
the scaling exponents are given by $\phi_n=(n+1-\tau)/(\tau-1)$ and
$\nu=(\tau-1)/2\sigma$. Here we present a study for $n=2$ because it is the lowest moment to diverge, but the results hold also for higher moments.

We  have  carried  out  a  scaling analysis  on  packings  created  by
molecular  dynamics (MD)  simulations (also  called  discrete elements
methods)  of   a  system  of   polydisperse  spheres  in   a  periodic
two-dimensional cell  subject to an isotropic  pressure. The particles
are deformable  and interact  via nonlinear Hertz-Mindlin  forces (see
Methods: Molecular Dynamics).  We first consider strongly polydisperse
frictionless packings at low  pressure, leading to strongly disordered
force   networks  with   clearly   visible  force   chains  shown   in
Fig.~\ref{fig1}~a.   For  different system  sizes,  we determined  the
force-chain clusters  at a large  number of threshold values  and thus
computed $m_2(f, N)$, the second moment of clusters sizes (leaving out
the largest  cluster in  each sample).  If  $m_2$ scales  according to
Eq.~(\ref{fss}),   then  plotting   $N^{-\phi}m_2$   as  function   of
$(f-f_c){N}^{1/2\nu}$ should lead to  a collapse of data for different
system sizes  on a  single curve. Varying  $\phi$, $\nu$ and  $f_c$, a
good collapse is indeed obtained for $\phi=0.89 \pm 0.01$ and $\nu=1.6
\pm 0.1$, clearly confirming scale invariance.

In order to test the dependence on various parameters, we repeated the
 procedure    for   packings    created   at    different   pressures,
 polydispersities,  coefficients of friction  and force-laws.   As the
 pressure is increased,  the grains deform and the  number of contacts
 per particle rises (see Fig. \ref{fig2} a). The force network becomes
 increasingly  uniform in  appearance, and  the distribution  of force
 magnitudes  narrower   (see  Supplementary  Information).   Computing
 $m_2(f)$ for  values of pressure spanning three  orders of magnitude,
 we nevertheless  find that the  optimal data collapse is  obtained in
 the  same   range  $\phi=0.89  \pm   0.01$  and  $\nu=1.6   \pm  0.1$
 independently of  the pressure.  Decreasing  the polydispersity leads
 to  crystallization of the  grains (see  Fig. \ref{fig2}  b), however
 such ordering  of the contact network  also leaves the  values of the
 scaling  exponents unchanged.  Moreover, although  friction  leads to
 smaller coordination number in the packing, it does not influence the
 scaling properties.   In this case the  forces are not  normal to the
 grains, but  tangential forces are typically much  smaller then $f_c$
 (see  Supplementary Information),  so  that they  do  not affect  the
 scaling  of  cluster  sizes.   Finally,  particles  interacting  with
 harmonic forces instead of non-linear Hertz forces lead once again to
 the same exponents.

\begin{figure}
\begin{center} 
\includegraphics[width=0.75\linewidth]{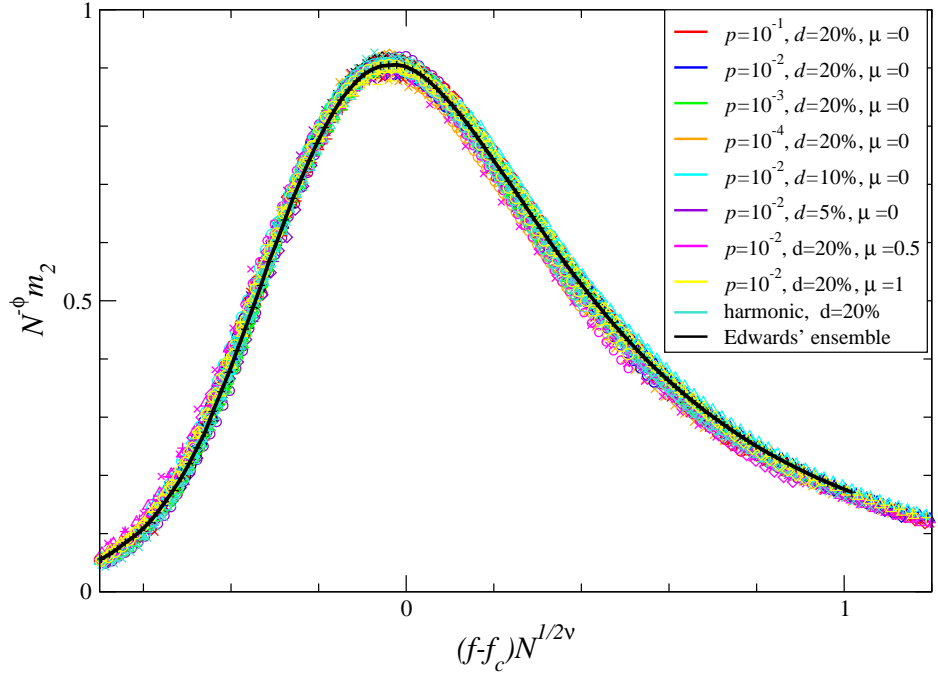}
\caption{The  universal  scaling  function.  The mean  square  of  the
  cluster sizes,  omitting the largest cluster, $m_2$,  is rescaled as
  $BN^{-\phi}m_2$ and  plotted as  function of the  rescaled threshold
  $A(f-f_c)N^{1/2\nu}$.  The figure  shows  the collapse  on a  single
  curve of 50 datasets corresponding to various packing parameters and
  system sizes.  A color is associated with each of six different sets
  of  parameters  (pressure  $p$,   polydispersity  $d$)  used  in  MD
  simulations.   For  each  set   of  parameters,  five  system  sizes
  (increasing by  factors of two)  are represented by  symbols $\circ,
  \diamond,  \triangle, \times,  +$.   The collapse  was obtained  for
  $\phi=0.89\pm0.1$  and $\nu=1.6\pm0.1$ for  all datasets,  while the
  value of $f_c, A$ and $B$  depends on the parameters (but not on the
  system size).  In black we  show Monte Carlo results of the Edwards'
  ensemble on the ``snooker packing'',  which gives the same values of
  $\phi$ and $\nu$.\label{fig3}}
\end{center}
\end{figure}

 The  independence  of  the  exponents  on  pressure,  polydispersity,
 friction and  force law  is clear evidence  of universality.   On the
 other hand,  as expected the value  of the critical  threshold is not
 universal, it varies between $1.3$ and $1.6$ times the average normal
 force,  and depends  on all  the parameters.   Moreover the  width and
 height of  the scaling function  vary with the parameters:  the width
 decreases with  increasing pressure, while the  height increases with
 increasing   polydispersity.   A  linear   rescaling  of   both  axes
 nevertheless leads to a single  collapse for all considered values of
 parameters, displayed  in Fig.  \ref{fig3}:  similarly to equilibrium
 critical  phenomena,   the  full  scaling  function   appears  to  be
 universal.

As the scaling exponents and  the scaling function are universal, they
can be obtained from simplified models in the appropriate universality
class.  Years ago, Edwards  proposed to generalize the micro canonical
principle of thermodynamics to jammed systems such as granular matter
\cite{edwards1}.   The key  idea is  to  ignore history  and treat  as
equally likely all stable  configurations of grains.  The approach was
found   to   be  successful   in   slowly   flowing  granular   matter
\cite{jam:makse1}, but so far  a quantitative confirmation in the case
of static granular matter has been lacking.

A simple  extension of this  micro-canonical ensemble to a  packing of
grains  in  a  fixed   geometry  was  named  the  ``snooker  packing''
\cite{fluct:jacco1}.  It consists of a hexagonal arrangement of rigid,
frictionless  and mono-disperse spheres  confined within  a triangular
domain, with  the same pressure applied  to all sides  of the packing.
In such  a system, force  balance alone does not  completely determine
the  forces. Here  it is  natural to  generalize Edwards'  approach by
treating  as equally likely  all arrangements  of repulsive  forces in
balance on each  grain, while keeping fixed the  geometry of contacts.
Using Monte Carlo simulations (see Methods: Monte Carlo), we generated
force networks  on the snooker  packing with uniform  probability, and
examined the scaling of the second moment of cluster size distribution
for  different system  sizes.  The  optimal collapse  is  obtained for
values of  scaling exponents  within error bars  equal to  those found
from MD  simulations.  Moreover, the full scaling  function appears to
be  identical  to  the  one  found  in  MD  simulations  as  shown  in
Fig. \ref{fig3}.

Some other simple models of  granular matter lead to different scaling
exponents.    For   example,   the   extensively   studied   $q$-model
\cite{qmodel0, qmodel, raj}, which implements balance between vertical
forces  only,  leads  to  $\phi=0.69\pm 0.01$  and  $\nu=3.1\pm  0.1$.
Moreover, the  universality class  depends on whether  a top-to-bottom
propagation  of  forces  is  assumed  (S.O  and  B.N.,  manuscript  in
preparation).   The agreement  between  the Edwards  ensemble and  the
results of  molecular dynamics simulations is  thus rather remarkable.
In  particular it  shows that  the elasticity  of the  grains  and the
details of the force it generates  are irrelevant for the shape of the
scaling  function.  More surprisingly,  the geometrical  randomness of
the  grain arrangements  does not  have any  bearing on  the universal
mechanical  properties.  We expect  that the  key ingredients  are the
vectorial balance of forces on each  grain, as well as the isotropy of
the  confining  pressure: static  packings  under  shear  may lead  to
different scaling  exponents.  We conjecture that  in three dimensions
the  exponents  and scaling  function  are  equally universal,  though
different from their two-dimensional equivalent.

Our results establish an unexpected connection between static granular
matter  and  equilibrium   critical  phenomena.   The  observed  scale
invariance is  in all  aspects comparable to  equilibrium criticality,
and defines  a novel universality  class of isotropic  force networks.
This  universality  class  is   for  example  distinct  from  that  of
percolation,  where the  same  analysis can  be  applied. This  result
implies  the existence  of long-range  correlations between  forces, a
characteristic  of structures such  as force  chains. We  expect force
networks  in other  jammed systems,  such as  foams and  emulsions, to
belong to the  same universality class.  We hope  that recent progress
in   the   measurement   of   inter-grain  and   inter-bubble   forces
\cite{majmudar05, fluct:brujic2} will lead
to an experimental evaluation of the scaling exponents.

\begin{methods}

\subsection{Molecular dynamics.}
We use molecular dynamics  simulations (also known as discrete element
method\cite{cundall79} in the engineering literature) to create static
granular packings  which mimic real granular matter.   Starting from a
dilute  gas phase,  we solve  Newton's equations  where  the particles
interact  with  a frictionless  repulsive  contact  force, called  the
Hertzian  interaction\cite{johnson85}.  The  centers of  the particles
are  restricted  to a  two-dimensional  plane;  however, the  Hertzian
repulsive  force   is  that  of   three-dimensional  elastic  spheres:
$F=\frac{2}{3}E/(1-\nu^2) R^{1/2}n^{3/2}$, where $E$ and $\nu$ are the
Young  modulus  and Poisson  ratio  of  the  particle's material,  and
$R=R_1R_2/(R_1+R_2)$.   The overlap  $n=R_1+R_2-r_{12}$  is calculated
from the particle's radii $R_1$  and $R_2$ and their centers' distance
$r_{12}$.   In   some  of  the   analysis,  we  used   the  frictional
Hertz-Mindlin\cite{johnson85}  force  law.   During the  evolution  we
shrink  the volume  of  the periodic  box  with rate  controlled by  a
feedback  loop to  achieve  a  target pressure.   As  we include  some
dissipation in the particle  interactions, eventually all motion stops
and a  static granular packing is obtained  in mechanical equilibrium.
For each set  of parameters we use 100  independent packings of 10,000
particles  in  two dimensions,  with  polydispersity ranging  5\%-20\%
(maximum  deviation   from  the  mean,  using   flat  distribution  of
radii). By  extracting recursively subsystems with half  the number of
grains, in total five system sizes were obtained.   The pressures are
expressed   in  units  of   the  Young   modulus  of   the  particle's
material. For more details see ref.~\onlinecite{somfai05}.

\subsection{Monte Carlo.}
The  Edwards ensemble  for a  granular heap  is a  uniform probability
measure on  the forces  subject to equations  for the balance  of each
particle and  inequalities for the  positivity of each  contact force.
In an appropriately reduced space of forces this represents a constant
non-zero  probability density  inside a  convex polyhedron.  We sample
this distribution by the  following stochastic process.  Starting with
a point inside  the polyhedron we select a line  in a random direction
(from a  discrete set of  directions). In this direction  we calculate
the intersections  with the polyhedron. A  new point is  chosen on the
line    with   uniform    probability    between   the    intersection
points. Averages calculated from such a process converge to unweighted
averages over the whole polyhedron. For more details see
ref.~\onlinecite{resp:epl}.
\end{methods}


\bibliography{gran_stat}


\begin{addendum}
 \item We thank B.~ Behringer, W.~Ellenbroek, M.~van Hecke,
    C.~Goldenberg, W.~van Saarloos and K.~Shundyak for discussions. SO is financially supported by the Dutch research organization FOM (Fundamenteel Onderzoek der Materie).
    ES is supported by the PHYNECS training network of the
    European Commission. 
 \item[Competing Interests] The authors declare that they have no
    competing financial interests.
 \item[Correspondence] Correspondence and requests for materials
    should be addressed to B.N.~(email: nienhuis@science.uva.nl).
\end{addendum}

\begin{figure*}
  \begin{center}
    \includegraphics[width=0.95\linewidth]{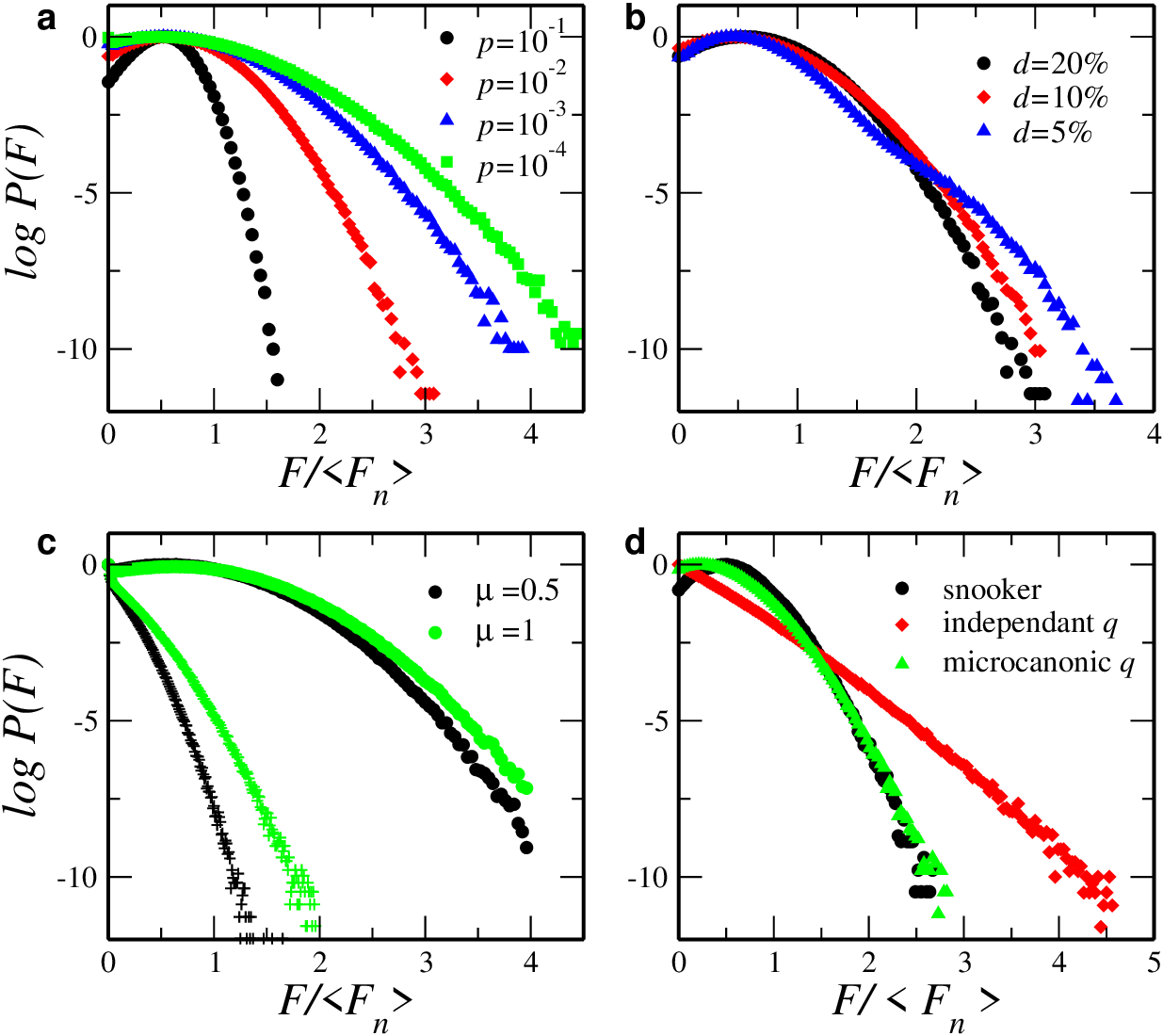}
    \caption*{{\bf Supplementary Figure:} Probability distributions $P(F)$ of force magnitudes in
    all the
    systems studied. All the forces are rescaled by the mean normal
    force $<F_n>$. (a-c), Dependence of $P(F)$ on the
    parameters varied in the molecular dynamics simulations: (a)
    pressure, (b) polydispersity and (c) friction. In (c), the
    distribution of normal forces is represented by the symbols
    $\bullet$, and the tangential forces by $+$. (d), $P(F)$ in the
    three
    lattice models mentioned in the text. }
  \end{center}
\end{figure*}

\end{document}